\documentclass[aps,prl,twocolumn,longbibliography,groupedaddress,floatfix,showpacs,reprint,superscriptaddress]{revtex4-2}
\usepackage{graphicx}
\usepackage{amsmath}
\usepackage{amssymb}
\usepackage{xcolor}
\usepackage{orcidlink}
\usepackage{gensymb}

\usepackage[T1]{fontenc}
\usepackage[utf8]{inputenc}
\usepackage{physics}
\usepackage{txfonts}
\usepackage{enumitem}
\usepackage{gensymb}
\usepackage{hyperref}
\usepackage{natbib}

\begin{document}
    \title{Creation and precise spectroscopy of $^{86}$Sr$_2$ halo molecules}
    
    \author{B. Iritani\orcidlink{0000-0002-7911-2755}}
        \affiliation{Department of Physics, Columbia University, 538 West 120th Street, New York, NY 10027-5255, USA}
    \author{J. Huang\orcidlink{0009-0005-3062-2972}}
        \affiliation{Department of Physics, Columbia University, 538 West 120th Street, New York, NY 10027-5255, USA}
    \author{W. Xu\orcidlink{0009-0005-4578-6785}}
        \affiliation{Department of Physics, Columbia University, 538 West 120th Street, New York, NY 10027-5255, USA}
    \author{G. Sim\orcidlink{0009-0005-9717-1950}}
        \affiliation{Department of Physics, Columbia University, 538 West 120th Street, New York, NY 10027-5255, USA}
    \author{R. Moszynski\orcidlink{0009-0008-7669-3751}}
        \affiliation{Quantum Chemistry Laboratory, Department of Chemistry, University of Warsaw, Pasteura 1, 02-093 Warsaw, Poland}
    \author{T. Zelevinsky\orcidlink{0000-0003-3682-4901}}
        \email{tanya.zelevinsky@columbia.edu}
        \affiliation{Department of Physics, Columbia University, 538 West 120th Street, New York, NY 10027-5255, USA}
    
	\date{\today}
\begin{abstract}
We report on the creation of $^{86}$Sr$_2$ molecules in the halo state and neighboring weakly bound states. Efficient molecule production via one-photon photoassociation relies on sufficient wavefunction overlap between the target vibrational states in the electronic excited-  and ground-state potentials. Using Autler-Townes spectroscopy, transition strengths are measured to identify optimal pathways
for production of weakly bound molecules. Vibrational splittings for the three least-bound vibrational states are measured, and dominant systematic uncertainties are evaluated with uncertainties below 100 Hz. From these splittings, absolute binding energies for these weakly bound vibrational states are determined. The results pave the way to a molecular isotope shift measurement with Sr$_2$.

\end{abstract}
\maketitle


The rovibrational structure of cold diatomic molecules provides special advantages over atoms in precision measurement, quantum simulation, and quantum computation~\cite{leungTerahertzVibrationalMolecular2023,acmecollaborationImprovedLimitElectric2018,roussyImprovedBoundElectrons2023,YeCarrollScience25_SpinDynamics,baoRamanSidebandCooling2023_PRX,demilleQuantumSensingMetrology2024}.
In particular, homonuclear alkaline-earth-metal dimers such as Sr$_2$ are well suited for both metrology and tests of fundamental physics due to their narrow optical and vibrational transitions and insensitivity to external fields~\cite{leungTerahertzVibrationalMolecular2023,kondovMolecularLatticeClock2019,tiberiSearchingNewFundamental2024}. 

The apparent incompatibility of general relativity and the Standard Model represents one of the most prominent issues in modern physics. Attempts to reconcile the two can result in a prediction of a new mass-dependent, non-Newtonian ``fifth force'' \cite{fayetNewInteractionsStandard1996,adelbergerTESTSGRAVITATIONALINVERSESQUARE2003,adelbergerTorsionBalanceExperiments2009,newmanTestsGravitationalInverse2009,knapenLightDarkMatter2017,kamiyaConstraintsNewGravitylike2015,NesvizhevskyPRD08_NeutronScatt,HeacockScience21_NeutronFifthForce}. The vibrational levels in Sr$_2$, dependent on the interactions between the constituent atoms, are a high-resolution nm-scale force sensor. Comparison between precise experimental measurements of these vibrational levels and quantum chemistry calculations that incorporate all known interatomic interactions can constrain these fifth forces \cite{tiberiSearchingNewFundamental2024}. Measurements of isotope shifts, rather than absolute binding energies for the complete molecular potential, can reduce the theoretical input to isotope-dependent terms.  The molecular structure of Sr$_2$ that lends itself to quantum chemistry calculations and the relative abundance of stable isotopes make it a good candidate for molecular vibrational isotope shift measurements to constrain these new interactions. $^{86}$Sr$_2$ has garnered special interest due to the existence of its weakly bound ``halo'' state, which has a very large spatial extent~\cite{jensenStructureReactionsQuantum2004, amanPhotoassociativeSpectroscopyHalo2018}. Previously, the $^{86}$Sr$_2$ halo state binding energy was measured using photoassociative spectroscopy of atoms, with the goal of quantifying the atomic scattering length and potentially studying Efimov physics via optical manipulation of interatomic interactions~\cite{mickelsonSpectroscopicDeterminationWave2005,amanPhotoassociativeSpectroscopyHalo2018,konHighintensityTwofrequencyPhotoassociation2019,borkowskiMassScalingNonadiabatic2014,jonesUltracoldPhotoassociationSpectroscopy2006,efimovEnergyLevelsArising1970,efimovEnergyLevelsThree1973,naidonEfimovPhysicsReview2017,kraemerEvidenceEfimovQuantum2006,knoopObservationEfimovResonance2009}.

Previously, we measured the systematic uncertainty of a $^{88}$Sr$_2$ vibrational transition spanning the ground state potential  to $4.6\times10^{-14}$~\cite{leungTerahertzVibrationalMolecular2023}. Aside from serving as a competitive secondary frequency standard in the THz frequency range \cite{RiehleMtrologia18_FreqencyStandards}, this measurement demonstrated the accessible precision and accuracy in searches for new forces. In combination with quantum chemistry theory, this level of precision can allow placing competitive laboratory constraints on a mass-dependent Yukawa force in the 0.1-1 nm range by measuring the isotope shifts of vibrational levels between $^{86}$Sr$_2$ and $^{88}$Sr$_2$~\cite{tiberiSearchingNewFundamental2024}. In the future, the analysis could be strengthened by utilizing $^{84}$Sr$_2$ molecules which have been observed previously~\cite{stellmerCreationUltracoldSr2012,ciameiEfficientProductionLonglived2017}.

\begin{figure}[h!]
\includegraphics[width=0.49\textwidth]{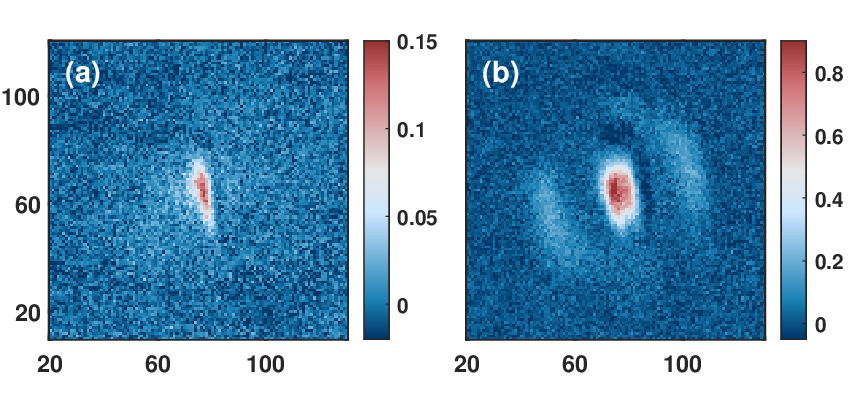}
\caption{Observation of $^{86}$Sr$_2$ molecule samples via photodissociation and subsequent absorption imaging of the resultant atomic fragments. Images are captured slightly off-axis from the optical lattice. Horizontal axis is along the direction of gravity, and both axes are in units of $\mu$m. Color bars represent optical density in arbitrary units. (a) The halo state, $v=-1$.  Only the state with total angular momentum $J=0$ is bound.  (b) The $v=-2$ molecules, with $J = 0$ dissociation at the center and $J = 2$ molecules receiving a greater kinetic energy and forming the outer ring; odd values of $J$ are forbidden by bosonic quantum statistics.
\label{fig:halo_figs}}
\end{figure}
Here, we spectroscopically investigate potential pathways to create weakly bound $^{86}$Sr$_2$ dimers via one-photon photoassociation (PA). We measure transition strengths for pairs of weakly bound states in the ground and lowest excited state potentials using photoassociative Autler-Townes spectroscopy.  Furthermore, we create samples of stable halo molecules (Fig.~\ref{fig:halo_figs}) in the threshold vibrational state $v=-1$ and in an absence of external fields.  We measure the vibrational energy intervals to neighboring weakly bound states and determine the dominant systematic uncertainties, lattice and probe induced Stark shifts, to <100 Hz uncertainty. Using the previously measured binding energy of the halo state \cite{amanPhotoassociativeSpectroscopyHalo2018}, we report the absolute binding energies for the weakly bound vibrational states $v=-2$ and $v=-3$. This work will inform high-precision ab initio quantum chemistry calculations for the ground-state potential of bosonic Sr$_2$, including relativistic and quantum electrodynamic (QED) effects that should be particularly pronounced at large bond lengths.
The results also pave the way for an isotope shift measurement between $^{86}$Sr$_2$ and $^{88}$Sr$_2$ to constrain non-Newtonian gravity or mass-dependent fifth forces.

An ultracold $^{86}$Sr sample is prepared by
cooling and trapping the atoms from a Zeeman-slowed beam with a two-stage magneto-optical trap (MOT) on the broad $^1 S _0$-$^1P_1$ transition and the narrow $^1S_0$-$^3P_1$  intercombination line. Subsequently, the atoms are transferred to a 914 nm one-dimensional optical lattice. After applying the spectroscopy pulse (or the photodissociation pulse, as in Fig. \ref{fig:halo_figs}), the remaining atoms are absorption imaged on the $^1S_0$-$^1P_1$ transition.
The narrow-line MOT laser is locked to a high finesse cavity via a Pound-Drever-Hall scheme and serves as the reference for phase locking the PA spectroscopy lasers \cite{leungTerahertzVibrationalMolecular2023,reinaudiOpticalProductionStable2012}.

\begin{figure}[t!]
\includegraphics[width=0.49\textwidth]{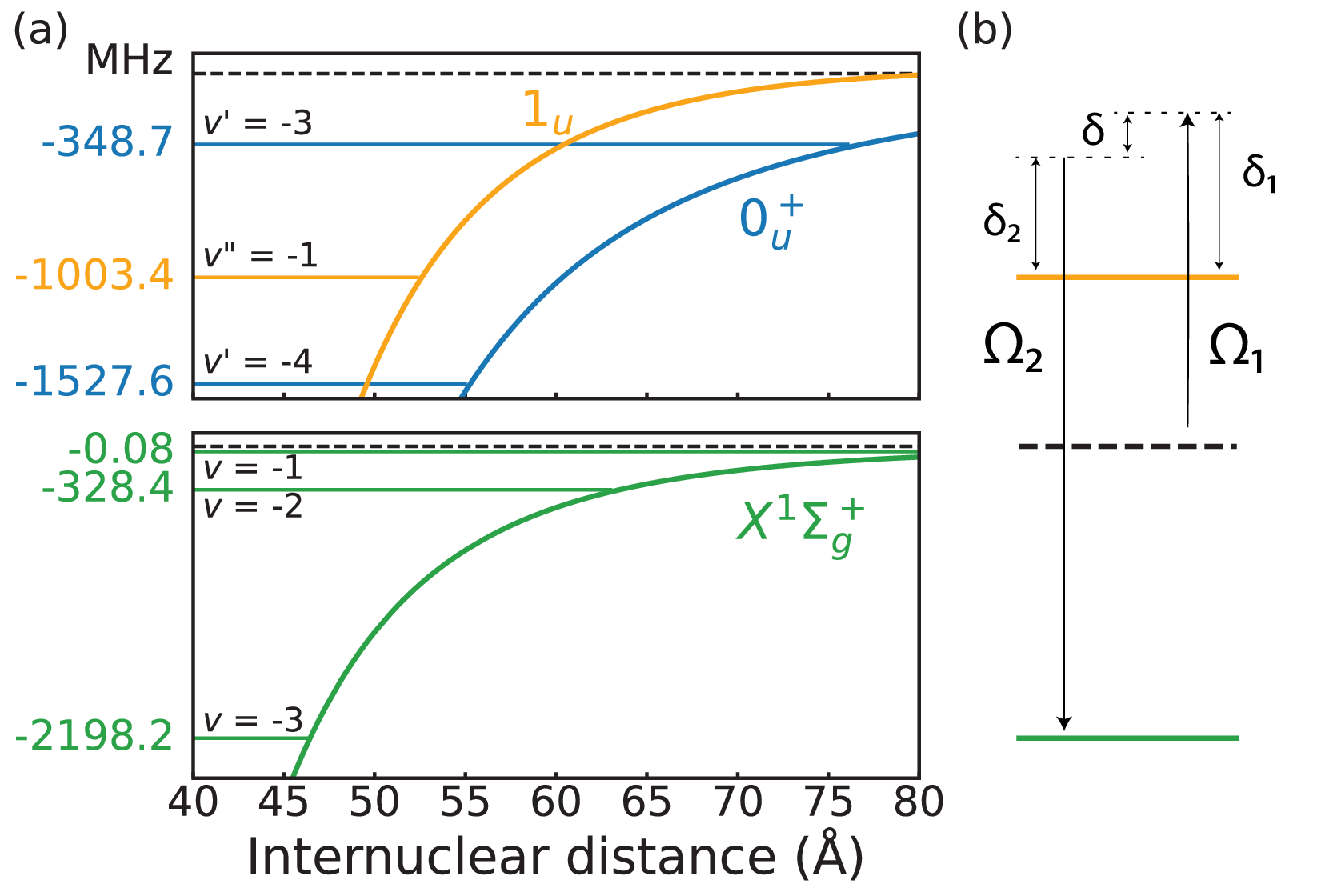}
\caption{(a) The ground-state electronic potential $X^{1}\Sigma_g^+$ asymptotes to the $^{1} S _{0}$+$^{1} S _{0}$ atomic threshold, and the excited-state potentials $1_{u}$ and $0_{u}^{+}$ asymptote to the $^{1} S _{0}$+$^{3} P _{1}$ intercombination threshold. The weakly bound vibrational states
are displayed ($v$, $v'$, and $v''$ for the three potentials, as labeled). (b) The Rabi frequency ($\Omega_1$, $\Omega_2$) and detuning ($\delta_1$, $\delta_2$, $\delta$) scheme for
photoassociative Autler-Townes spectroscopy.
\label{fig:potentials_fig}}
\end{figure}

Our PA method involves one-photon excitation of ground state atoms into weakly bound vibrational states in the $0_u^+$ or $1_u$ potential near the $^1S_0$-$^3P_1$ atomic threshold (Fig. \ref{fig:potentials_fig}(a)). The molecules then decay into several weakly bound states of the ground state electronic potential, $X^1\Sigma_g^+$. The branching ratios for this decay are governed by the relative transition strengths. The near-threshold transition strength is proportional to the vibrational wavefunction overlap, or the Franck-Condon factor.  A near-unity Franck-Condon factor ensures efficient creation of molecules in a single vibrational state $v$ of the ground state \cite{reinaudiOpticalProductionStable2012}.

\begin{figure}[h!]
\includegraphics[width=0.49\textwidth]{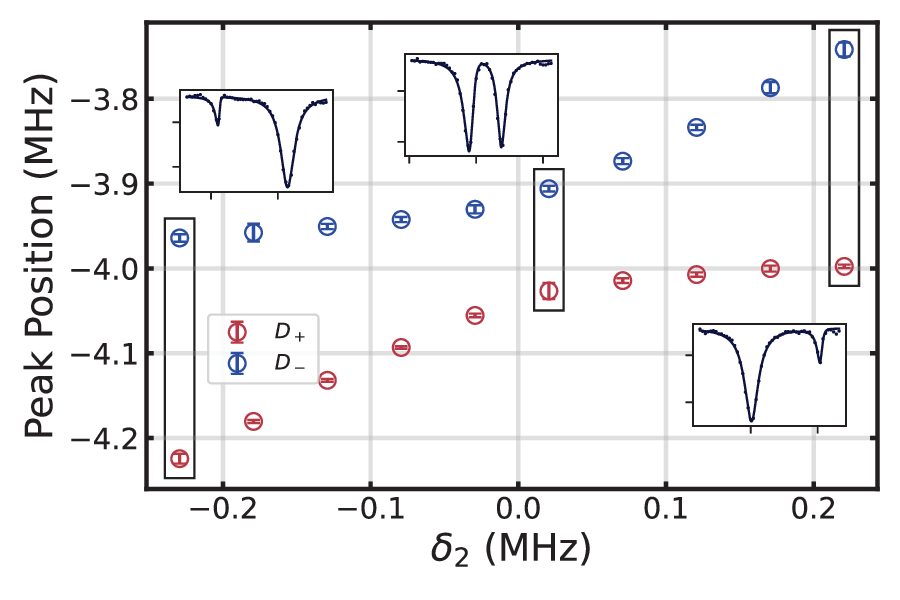}
\caption{Autler-Townes doublet peak positions for the $1_u(-1)$ and $X(-3)$ pair of states. Example scans are shown in insets.
\label{fig:Autler_Town_egs}}
\end{figure}
To measure the transition strengths, we use photoassociative Autler-Townes spectroscopy. In this two-photon process, the pump laser (Rabi frequency $\Omega_1$) couples free atoms to the excited intermediate state, and the anti-Stokes laser (Rabi frequency $\Omega_2$) connects the intermediate state to the target state. The anti-Stokes laser is kept at a fixed frequency while the pump laser is scanned to reveal the characteristic Autler-Townes splitting of the photoassociation peak. The detuning configuration is shown in Fig. \ref{fig:potentials_fig}(b).  For each measurement, we fit an Autler-Townes doublet of the form~\cite{leungTransitionStrengthMeasurements2020,LeungThesis}
\begin{equation}
\begin{split}
N(\delta_1, & \delta_2, t)= N_0 \exp \Bigg[-2\pi t \frac{\Gamma\left|\Omega_1\right|^2}{\Gamma^2+4 \delta_1^2} \\
&\times\left(1-\left|\Omega_2\right|^2 \frac{\left|\Omega_2\right|^2-8 (\delta_1-\delta_2) \delta_1+\Gamma_{\mathrm{eff}} \Gamma\left(1-4 \delta_1^2 / \Gamma^2\right)}{\left|\Omega_2\right|^2+\left|\left(\Gamma+2 i \delta_1\right)\left(\Gamma_{\mathrm{eff}}+2 i (\delta_1-\delta_2)\right)\right|^2}\right)\Bigg]
\end{split}
\label{eq:AT_shape}
\end{equation}
where $\delta_1$ and $\delta_2$ are the detunings of the pump and anti-Stokes lasers, and $t$ is the probe time.
$\delta_2$ are the detunings of the pump and anti-Stokes beams respectively. $\Omega_1$ and $\Omega_2$ are the corresponding Rabi frequencies.
$\Gamma$ is the excited state linewidth, dominated by power broadening, and $\Gamma_{\rm{eff}}$ is a decoherence rate that represents the relative linewidth between the pump and anti-Stokes lasers. $\Gamma_{\rm{eff}}$ is a free parameter that was fit to each scan, with an average value of 14(4) kHz.

If only the pump laser is present, i.e., the one-photon free-to-bound transition is driven, then we can set $\Omega_2=\delta_2=\Gamma_{\rm{eff}}=0$ and Eq. (\ref{eq:AT_shape}) becomes
\begin{equation}
N = N_0\exp(\frac{-2\pi t|\Omega_1|^2\Gamma}{\Gamma^2+4\delta_1^2}).
\label{eq:TA_shape}
\end{equation}
For each excited state, $t$ is fixed and $\delta_1$ is scanned to locate the resonance. Next, the laser is kept on resonance while $t$ is varied, in which case Eq. (\ref{eq:TA_shape}) becomes $N=N_0\exp(-2\pi t\Omega_1/\Gamma
)$. The parameters $\Omega_1$ and $\Gamma$ are found by fitting these decay curves.
Finally, with both lasers turned on we obtain a series of Autler-Townes spectra as shown in the insets of Fig. \ref{fig:Autler_Town_egs}. 

\begin{figure*}[]
\centering
  \includegraphics[width=1\textwidth]{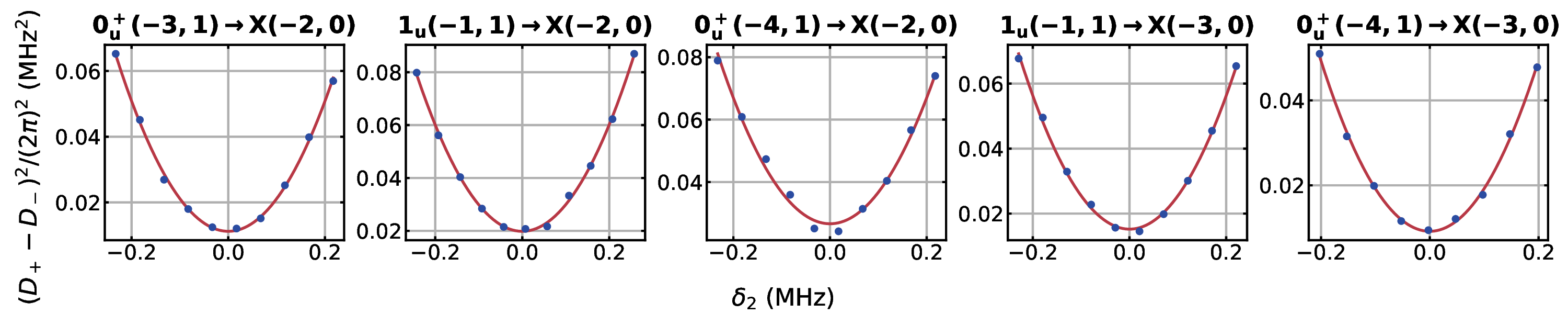}

  \caption{All Autler-Townes doublet separation parabolas. The minima of the parabolas are proportional to the relative transition strengths. The quantum numbers in parentheses correspond to ($v,J$).
  \label{fig:TransitionStrength_parab}}
\end{figure*}

From fitting the spectral lines to Eq. (\ref{eq:AT_shape}), we can directly determine $\Omega_2$ and therefore the relative anti-Stokes transition strengths.  Alternatively, the splitting between the higher- and lower-frequency Autler-Townes peaks depends on $\delta_2$ as
\begin{equation}
 \left(D_+-D_-\right)^2 = \delta_2^2 + \left|\Omega_2\right|^2.
 \label{eq:Split}
\end{equation}
Therefore, finding the minimum of this parabola, measured with constant laser beam intensities, yields $|\Omega_2|^2$ and the relative transition strengths.
Measuring the parabolas for several pairs of vibrational states yields a direct comparison of transition strengths. The results for all 5 measured state pairs with sizable transition strengths are displayed in Fig. \ref{fig:TransitionStrength_parab} and Table \ref{tab:TStrengths}. The quoted transition strengths are normalized to the largest measured value.


\begin{table}[b!]
\begin{tabular}{|c||c|c|c|}
    \hline \rule{0pt}{10pt} $ { }^{1}\mathrm{S}_0+{ }^3 \mathrm{P}_1$ levels  & \multicolumn{2}{|c|}  {${ }^1 \mathrm{S}_0+{ }^1 \mathrm{S}_0$ levels} \\
       & $v=-2$ & $v=-3$ \\
    \hline $0_u^+(-3,1)$ &  {0.42(2)}& -\\
    \hline $1_u(-1,1)$&  
       {0.74(2)}  &  {0.57(2)}\\
    \hline $0_u^+(-4,1)$  &  {1.00(4)}  &  {0.34(2)}\\
    \hline
\end{tabular}
  \caption{Summary of measured relative transition strengths for photoassociation pathways.  The values are normalized to the strongest transition.}
  \label{tab:TStrengths}
\end{table}


Next we proceed to creating and detecting molecules. Guided by the transition strength measurements summarized in Fig. \ref{fig:TransitionStrength_parab}, we photoassociate ground state atoms to $0_u^+(-4,1)$
and allow them to decay into the $X(-2,0)$ state. An optional two-photon Raman $\pi$-pulse is then applied via the $0_u^+(-2,1)$ ($-44.1$ MHz) state to move the molecules to the halo state $v=-1$, and a resonance signal is observed as loss from $v=-2$. At the end of each sequence, the remaining $v=-2$ molecules are detected by first photodissociating to the $^1S_0+{^3P}_1$ threshold and then absorption imaging the
atoms.
We also can directly photodissociate the halo state to the $^1S_0+{^3P}_1$ threshold (Fig. \ref{fig:halo_figs}(a)). Attempts to directly populate the $v=-1$ halo state with one-photon photoassociation produced smaller molecule samples.  The $1/e$ lifetimes of the $v=-2$ and $v=-1$ halo molecule samples were measured to be $\sim50$ ms, limited by two-body collisional loss.

\begin{figure}[b!]
\includegraphics[width=0.49\textwidth]{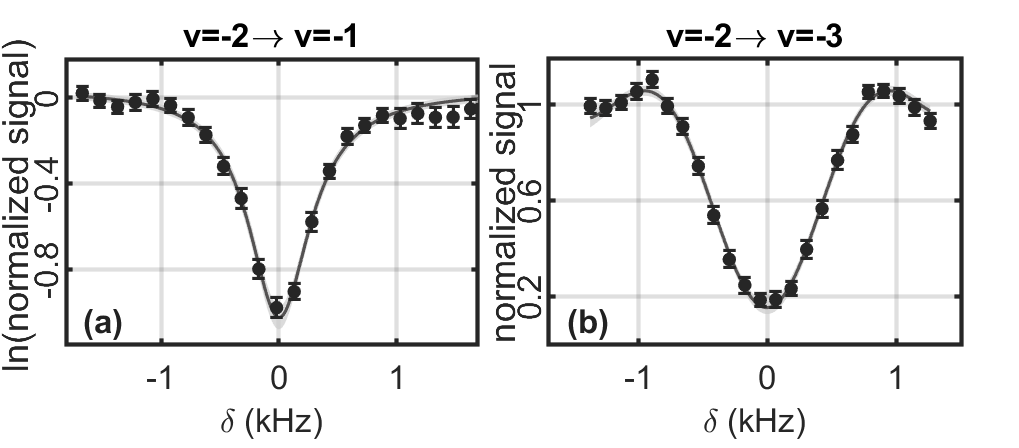}
\caption{Spectra of Raman transitions between weakly bound states at operational intensities for (a) $v=-2\rightarrow v=-1$ and (b) $v=-2\rightarrow v=-3$. Black points are (a) natural logarithm of normalized signal and (b) normalized signal versus $\delta$. Fitted curves have (a) Lorentzian and (b) Rabi lineshapes. The error bars display standard error of the mean.
\label{fig:rabiVSlortz}}
\end{figure}

In order to characterize the ultralong-range portion of the molecular potential, we measure energy splittings between the 3 least bound vibrational states. This is done using two-photon Raman spectroscopy of a trapped molecular sample, utilizing weakly bound states in the $0_u^+$ potential as intermediate states. Typical lineshapes are shown in Fig. \ref{fig:rabiVSlortz}. For each trace the anti-Stokes laser frequency is fixed, and the pump laser is scanned across the resonance. The common detuning is much larger than any residual Zeeman splitting in $0_u^+$. The vibrational splitting is the difference between the Raman laser frequencies. Both Raman lasers are phase-locked to the 689 nm narrow-line MOT laser, which itself is PDH locked to a high finesse cavity, with different offsets. An additional offset between the Raman laser frequencies arised from double-passed acousto-optic modulators (AOMs) for both lasers. RF frequencies for the offsets and AOMs are generated by direct digital synthesizers that are referenced to a commercial Rb frequency standard steered by a GPS signal.

The resonance position for the $v=-2\rightarrow v=-1$ transition is obtained by fitting a Lorentzian lineshape (Fig. \ref{fig:rabiVSlortz}(a)). Spectra for the $v=-2\rightarrow v=-3$ transition, are fit to a Rabi lineshape due to laser-induced power broadening,  (Fig. \ref{fig:rabiVSlortz}(b)),
\begin{equation}
N = \frac{|\Omega|^2}{|\Omega|^2+\delta^2} \sin ^2\left(\frac{\sqrt{|\Omega|^2+\delta^2}}{2}t\right),
\end{equation}
where $\Omega$ and $\delta$ are the overall Rabi frequency and detuning.

\begin{figure}[h!]
  \vspace{-1.3cm}
  \centering
  \includegraphics[width=0.49\textwidth]{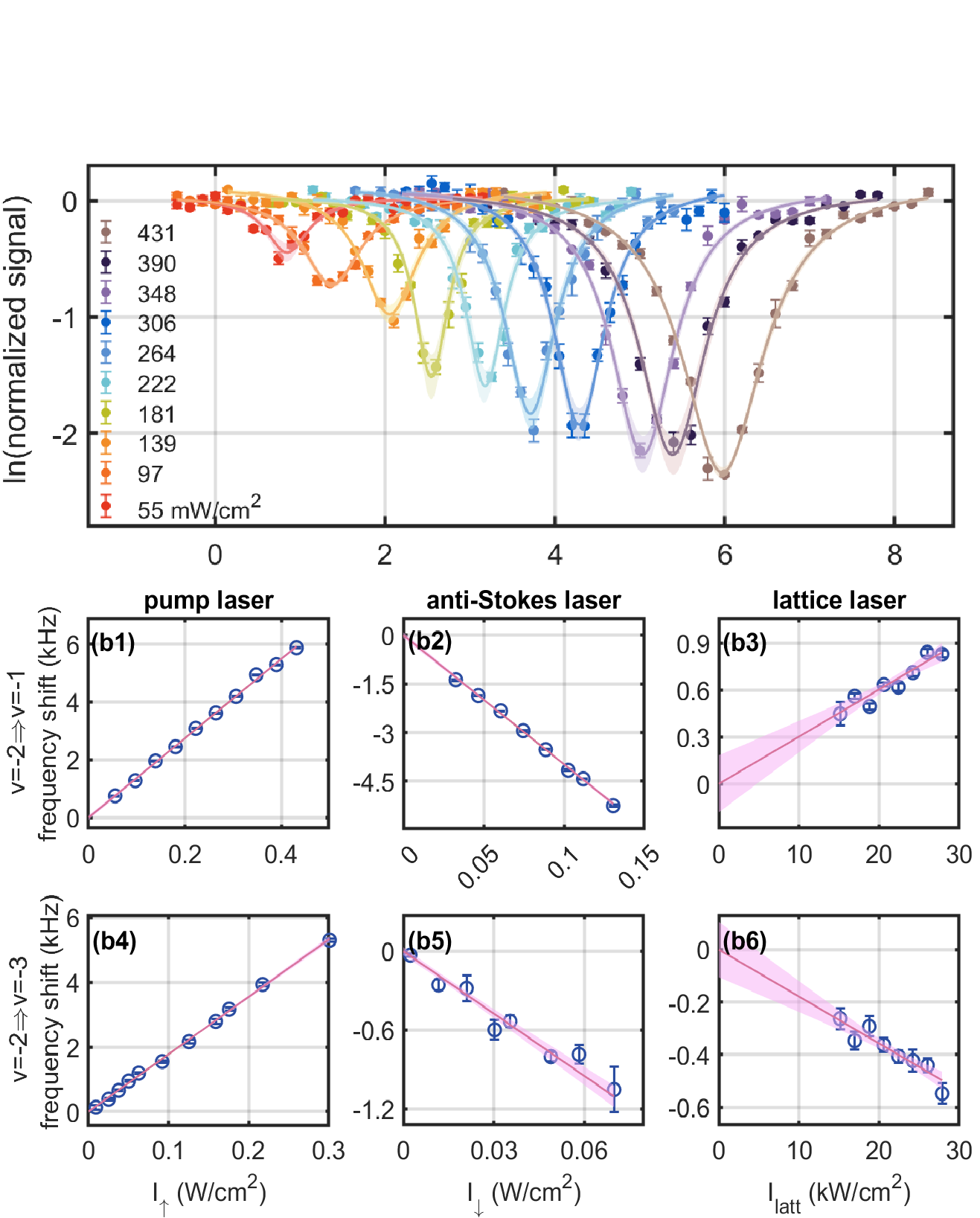}
  \caption{(a) Sample spectra of a two-photon Raman transition exhibiting probe light shifts. Points represent data for 10 different upleg probe laser intensities. Error bars are statistical standard errors obtained from 5 trials.
  Lines represent Lorentzian fits of the natural logarithm of the detected atom number.  Lower panels:  Shifts induced by probe and lattice lasers versus light intensity for $v=-2\rightarrow v=-1$ (b, top row) and $v=-2\rightarrow v=-3$ (b, bottom row). (b1,b4) Light shifts induced by the pump laser; (b2,b5) light shifts induced by the anti-Stokes laser; and (b3,b6) light shifts induced by the lattice laser.}
  \label{fig:lightshift_combined}
\end{figure}
\begin{table}[b!]
\begin{tabular}{|c||c|c|}
\hline Stark shift (\textbf{Hz}) & $v=-2\rightarrow v=-1$  &   $v=-2\rightarrow v=-3$    \\
\hline Pump & 1820$(50)$  & $742(32)$ \\
\hline Anti-Stokes & $ -1855(46)$ &  $-147(25)$ \\
\hline Lattice & $567(37)$  & $-335(23)$ \\
\hline \textbf{Total} & $ \mathbf{532(77)} $ & $\mathbf{259(47)}$\\
\hline 
  \end{tabular}
  \caption{Leading systematic shifts (light shifts due to the probe and lattice lasers) for $v=-2\rightarrow v=-1$ and $v=-2\rightarrow v=-3$ vibrational splitting measurements.}
  \label{tab:systematics}
\end{table}
By far the dominant systematic effects are the Stark shifts due to the probe and lattice lasers. We evaluate these systematics to <100 Hz uncertainty. For both transitions and for each of the contributing systematic effect, peak position are measured at several laser intensities, the range of which is limited by laser power availability (Fig. \ref{fig:lightshift_combined}). Five trials are used for each point, and a weighted linear fit is performed as a function of intensity. The shifts are then extrapolated to zero intensity. The systematic corrections and their uncertainties are reported in Table \ref{tab:systematics}.

\begin{table}[t!]
\begin{tabular}{|c||l|}
\hline Vibrational state & Binding energy (kHz)   \\
\hline $v = -1$ & $-83.00(7)(20)$  \cite{amanPhotoassociativeSpectroscopyHalo2018}  \\
\hline $v = -2$ & -328,477.288$(226)$   \\
\hline $v = -3$ & -2,198,158.333$(231)$   \\
\hline 
  \end{tabular}
  \caption{Binding energies of weakly bound states in the $X^1\Sigma _{g}^+$ potential of $^{86}$Sr$_2$, calculated with the measured vibrational splittings and the halo state binding energy previously determined via photoassociative spectroscopy \cite{amanPhotoassociativeSpectroscopyHalo2018}.}
  \label{tab:binding energies}
\end{table}
Accounting for the systematic shifts, we report the vibrational splittings. We have measured the $v=-2\rightarrow v=-1$ vibrational splitting to be 328.394 288 (78) MHz and the $v=-2\rightarrow v=-3$ splitting to be 1869.681 045 (47) MHz. The absolute binding energy of the $v=-1$ halo state was previously measured via photoassociative spectroscopy to be $-83.00(7)(20)$ kHz \cite{amanPhotoassociativeSpectroscopyHalo2018}. Therefore, we can report the absolute binding energies of the weakly bound states $v=-2$ and $v=-3$ (Table \ref{tab:binding energies}).


Nonpolar diatomic molecules provide a promising avenue for quantum chemistry, metrology, and precision tests of fundamental physics \cite{mcdonaldPhotodissociationUltracoldDiatomic2016,leungTerahertzVibrationalMolecular2023,tiberiSearchingNewFundamental2024,borkowskiProbingNonNewtonianGravity2017,borkowskiWeaklyBoundMolecules2019}. $^{86}$Sr$_2$ molecules are especially intriguing because they possess a weakly bound halo state which has an unusually large spatial extent without applied magnetic fields and can provide unique insights into atomic and molecular physics such as long-range QED in bound systems. In this work, we have identified favorable pathways for producing weakly bound $^{86}$Sr$_2$ molecules by measuring a range of transition strengths via two-photon Autler-Townes spectroscopy in a trapped $^{86}$Sr cloud.  We have subsequently created stable samples of $^{86}$Sr$_2$ molecules in the halo state and two neighboring bound states.  This is a crucial starting point for a molecular vibrational isotope shift measurement, for tests of relativistic effects in molecular systems and laboratory searches for new gravity-like forces in the 0.1-1 nm range \cite{tiberiSearchingNewFundamental2024}. Using the newly created molecular samples, we have measured vibrational splittings for the three most weakly bound states with $<100$ Hz uncertainties by evaluating the dominant systematic uncertainties arising from finite trapping and probing laser intensities. These results serve as key inputs for characterizing the halo state and ab initio quantum chemistry calculations.\\

We thank D. Mitra, M. Borkowski, and A. Wozniak for helpful discussions, and J. Dai and B. C. Riley for experimental assistance. We are grateful for financial support from NSF grant PHY-2409389, the Brown Science Foundation, and the Chu Family Foundation.


%

\end{document}